\documentclass [twocolumn,prb] {revtex4}
\usepackage {graphicx}
\usepackage {epsf}
\usepackage {amsmath} 
\usepackage {amssymb}
\usepackage {bbm}
\usepackage {latexsym}

\newcommand{\ddk}[1]{\frac{d^3 #1}{(2\pi)^3}}
\renewcommand{\vec}[1]{{\bf \boldsymbol #1}}

\def\F{{\mathcal F}}

\def\O{{\mathcal O}}
\def\T{{\mathcal T}}

\def\i{{\rm i}}

\def\G{\Gamma}

\def\b{\beta}

\def\d{\delta}
\def\e{{\rm e}}
\def\g{\gamma}

\def\th{\theta}
\def\w{\omega}

\begin{document}

\title{Field-tuned quantum critical point of antiferromagnetic metals}

\date{\today}

\author{I. Fischer}
\affiliation{Institut f\"ur Theoretische Physik, Universit\"at zu K\"oln, 50937 K\"oln}
\author{A. Rosch}
\affiliation{Institut f\"ur Theoretische Physik, Universit\"at zu K\"oln, 50937 K\"oln}

\begin{abstract}
  A magnetic field applied to a three-dimensional antiferromagnetic
  metal can destroy the long-range order and thereby induce a quantum
  critical point.  Such field-induced quantum critical behavior is the
  focus of many recent experiments. We investigate theoretically the
  quantum critical behavior of clean antiferromagnetic metals subject
  to a static, spatially uniform external magnetic field. The
  external field does not only suppress (or induce in some systems)
  antiferromagnetism but also influences the dynamics of the order
  parameter by inducing spin precession.  This leads to an exactly
  {\em marginal} correction to spin-fluctuation theory. We investigate
  how the interplay of precession and damping determines the specific
  heat, magnetization, magnetocaloric effect, susceptibility and
  scattering rates. We point out that the precession can change the
  sign of the leading $\sqrt{T}$ correction to the specific heat
  coefficient $c(T)/T$ and can induce a characteristic maximum in
  $c(T)/T$ for certain parameters. We argue that the susceptibility
  $\chi =\partial M/\partial B$ is the thermodynamic quantity which
  shows the most significant change upon approaching the quantum
  critical point and which gives experimental access to the
  (dangerously irrelevant) spin-spin interactions.
\end{abstract}

\maketitle

The study of quantum phase transitions is currently a very active
field of research in theoretical as well as experimental condensed
matter physics. Particularly in a large number of metals -- mostly
heavy Fermion or transition metal compounds -- the critical
fluctuations associated with a quantum phase transition induce
anomalous behavior in thermodynamic and transport quantities like
diverging specific heat coefficients or a linear resistivity quite
distinct from the behavior of a conventional Fermi liquid.

Experimentally there are three main methods to tune a system towards a
quantum critical point: doping, pressure, and magnetic field.  Doping
has the disadvantage that it induces disorder and it cannot be easily
adjusted within a single sample. These problems are absent if pressure
is used as the control parameter of the quantum phase transitions.
However, the presence of a pressure cell makes many experiments
difficult. For this reason, many recent experiments
\cite{heuser98,loehneysen01,bauer00,steglich,CeIrIn,sarrao,paglione,metamagneticQPT,tlcucl,srcubo,BaCuSi2O6}
investigate field-tuned quantum critical behavior, where an external
magnetic field is used to control the distance from the quantum
critical point.  Generally it is expected that the presence of a
magnetic field changes the universality class of the transition as in
its presence time reversal invariance is broken.  In this paper, we
will therefore analyze theoretically the quantum critical behavior of
a clean itinerant antiferromagnet in three dimensions subject to a
static, spatially uniform external magnetic field $B$.

Such a situation has been investigated in a number of
experiments\cite{heuser98,loehneysen01,bauer00,steglich,CeIrIn}.  For
example in CeCu$_{5.2}$Ag$_{0.8}$ \cite{heuser98} and
CeCu$_{5.8}$Au$_{0.2}$ \cite{loehneysen01} magnetic order can be
suppressed by moderate magnetic fields.  In these systems the quantum
critical behavior induced by a magnetic field $B$ appears to be
qualitatively different compared to the critical properties for
vanishing field (controlled by pressure or doping). In the
presence of a field these systems seem to follow\cite{heuser98} the
predictions from spin-fluctuation theory
\cite{hertz,millis,moriyaBuch} for three-dimensional nearly
antiferromagnetic metals, while this is not the case for $B=0$ \cite{CeCuAu}.  Similarly
experiments\cite{bauer00}
 in field tuned YbCu$_{5-x}$Al$_x$ appear to be consistent
with spin-fluctuation theory, which is not found to be the case in
YbRh$_2$Si$_2$ where magnetic order is suppressed by tiny magnetic
fields \cite{steglich}. Recently, in CeCoIn$_5$ \cite{sarrao,paglione}
the superconducting order was suppressed by a magnetic field -- it is
at the moment a controversial question whether the observed anomalous
behavior is related to a superconducting quantum critical point or
whether magnetism plays a role in this system.

Another interesting class of systems are {\em insulators} like
TlCuCl$_3$ \cite{tlcucl}, SrCu$_2$(BO$_3$)$_2$ \cite{srcubo} or
BaCuSi$_2$O$_6$ \cite{BaCuSi2O6} where antiferromagnetic order has
been {\em induced} by the application of a magnetic field $B$. These
transitions \cite{tlcucl} can be interpreted as a Bose-Einstein
condensation (see below) of spin-1 excitations. The energy of
the ``spin-up'' component of such triplets is lowered by $B$ until it
condenses at a critical field, $B=B_c$, thereby inducing
antiferromagnetic order perpendicular to the magnetic field.

In contrast to classical transitions, the dynamics, i.e.~the temporal
quantum fluctuations, of the order parameter determines the nature and
universality class of a quantum phase transition. For example, at the
critical point of an insulating antiferromagnet, the dynamics of the
order parameter $\Phi$ can be described \cite{sachdev} as in a Klein
Gordon equation $(\partial_t^2-\nabla^2) \Phi$. In such a system,
typical frequencies $\omega$ scale linearly with the momentum, $\omega
\propto q^z$, where $z=1$ is the dynamical critical exponent. In
contrast, in a metal the excitation of particle-hole pairs leads to a
Landau damping \cite{hertz} of the antiferromagnetic order parameter,
$(\partial_t+\nabla^2)\Phi$ and therefore $z=2$. Here we assumed that
the ordering vector $\vec{Q}$ is sufficiently small, $Q<2 k_F$, such
that low-energy particle-hole pairs with momentum $\vec{Q}$ exist.

\begin{figure}
\includegraphics*[width=0.75 \linewidth]{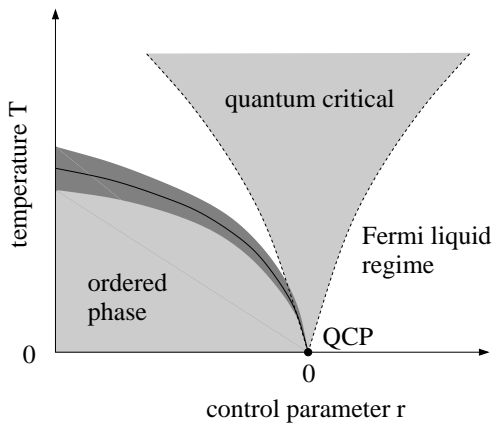}
\caption{Schematic phase diagram of a quantum phase transition with a control parameter $r \propto B-B_c$.\label{phaseDia}}
\end{figure}
A magnetic field will have two main effects: first it will suppress (or in some cases\cite{CeIrIn} also induce) magnetic order.
More interesting is the second effect: It induces a precession of the magnetic moments $\vec{S}$ perpendicular to the magnetic field
\begin{equation}\label{precession}
\partial_t \vec{S}= \vec{B} \times \vec{S}
\end{equation}
and therefore modifies the dynamics of the order parameter. The linear
time-derivative also translates to a dynamical exponent $z=2$, and
therefore the question arises how the precession competes with damping
in a metal, which is characterized by the same $z$. For
insulating systems the physics of the precession term has been widely
discussed \cite{sachdev,affleck,nikuni,giamarchi,bruce,haas}. The
corresponding quantum critical behavior as a function of magnetic
field of such an insulating magnet in an external field is actually
well known: it is expected to be in the same universality class as the
quantum phase transition of a low-density interacting Bose Einstein
condensate as a function of chemical potential. The linear time
derivative $i \partial_t \Psi$ of the Schr\"odinger equation can in
this case be identified with the precession term (\ref{precession}),
see below.

%When the external magnetic field applied to such a system has reached a certain strength, the magnons can form a Bose-Einstein condensate. Evidence for such a Bose-Einstein condensation of magnons has been reported in \cite{nikuni}. 

In the following, we study the interplay of ohmic damping and spin
precession terms in the case of a nearly antiferromagnetic
metal. First we present the model for the order parameter field and a
short derivation of the effective action.  Then we list the
renormalization group equations for the parameters of the model and
use them to derive the behavior of the correlation length. In the
following sections we calculate the specific heat, thermal expansion,
magnetocaloric effect, and susceptibility. We show for example that
sufficiently large magnetic fields can induce sign changes in the
critical contribution to the specific heat and that the susceptibility
is particularly suited to probe the vicinity of the quantum critical
point.  Finally, we investigate the influence of the $B$-field on the
scattering rate of the electrons.

\section{Model and Effective action}

Following Hertz\cite{hertz}, we describe the critical behavior of an
antiferromagnetic metal entirely in terms of an effective
Ginzburg-Landau-Wilson theory of an order parameter field
$\vec\Phi(\vec{r},t)$ which represents the fluctuating (staggered)
magnetization of the system.

In the absence of a magnetic field, the quadratic part of the action
takes the form\cite{hertz} (assuming negligible spin-orbit coupling)
\begin{equation}\label{damping}
  S_2^\prime[\vec \Phi] = \frac 1 \b \int \ddk{k} \sum_n
  \Phi^*(r + k^2 + |\w_n|) \Phi,
\end{equation}
where $r$ measures the distance from the quantum critical point and
momenta $\vec{k}$ are given relative to the antiferromagnetic ordering
wave vector $\vec{Q}$.  The $|\w_n|$ term arises from the
(Landau-) damping of the spin-fluctuations by gapless fermionic
excitations in the vicinity of points on the Fermi surface that are
connected by $\vec Q$ (assuming $Q<2 k_F$).

How will this effective action change in the presence of a magnetic
field?  First, $r=r(B)$ will acquire a magnetic field dependence,
for example, $r$ will grow for larger fields in systems where 
antiferromagnetism is suppressed by $B$. Second, the magnetic
field breaks the rotational invariance and components of $\Phi$
parallel and perpendicular to $\vec{B}$ will have different masses,
$r_z$ and $r_\perp$, respectively. Third, as argued above, the magnetization will precess around $\vec{B}$; this is described by an extra term
(in coordinate and time space for convenience)
\begin{eqnarray}\label{precession1}
  S_2^{pr}[\vec \Phi] &=& \int \limits_0^\b \! d\tau \int \! d{\bf r} \, \vec
  b \cdot \i (\vec \Phi \times \partial_t \vec \Phi) \\
&=& \int\!\! \int  b (\i \Phi_x
  \partial_t \Phi_y - \i \Phi_y \partial_t \Phi_x) =\int \!\! \int  b \tilde\Phi_\perp^* \partial_t \tilde\Phi_\perp \nonumber
\end{eqnarray}
in the effective action, where $\vec b$ is parallel to $\vec B$
(taken to point into the $\hat z$-direction) and we have introduced
the complex field $\tilde\Phi_\perp = \Phi_x + \i \Phi_y$. Note that
(\ref{precession1}) breaks time-reversal invariance. Therefore such a
term is absent for $\vec{B}=0$.

Above, we deduced the form of the effective action on phenomenological
grounds but it can also be derived from a more explicit calculation
starting from a Hubbard-type model
of electrons moving in the presence of a magnetic field,
$H=\sum_{\vec{k}\sigma} (\epsilon_{\vec{k}}+B \sigma^z_{\sigma\sigma})
\psi^\dagger_{\vec{k}\sigma}\psi_{\vec{k}\sigma}+U \sum n_\downarrow
n_\uparrow$. Here, the magnetic field enters only via a Zeeman-term;
we do not take orbital effects into account, assuming that Landau
levels are broadened by disorder or thermal effects. Note that in the
experimentally most relevant heavy Fermion system, orbital effects are
strongly suppressed compared to contributions from the Zeeman term as
the effective masses and magnetic susceptibilities are very large in
those systems\cite{rosch:2000}.

For simplicity, we assume that the antiferromagnet is commensurate
(incommensurate antiferromagnets show the same qualitative behavior
for all quantities discussed below) and introduce a real order
parameter vector $\vec \Phi(\vec x,t)$ as a Hubbard-Stratonovich field
which decouples the spin-density part of the interaction.  Following
Hertz, the electrons are now integrated out to obtain an effective
action for the order parameter, generating a priori infinitely many
interaction terms.  We truncate the effective action, retaining the
leading frequency and momentum dependence of the Gaussian part of the
action as well as a constant $\vec \Phi^4$ interaction term, since all
other terms are irrelevant in the renormalization group
sense\cite{hertz,millis} (cubic terms are discussed in appendix~\ref{cubic}). 
For the quadratic part one obtains $S_2=
\frac 1 \b \sum_{\omega_, \vec{k}} {\Phi}_{\omega_n,\vec{k}}^\alpha
(\delta_{\alpha \alpha'}/J + \chi^0_{\alpha \alpha'}(\vec{k},\i \w_n))
\Phi^{\alpha'}_{-\omega_n,-\vec{k}}$ where $J$ is the interaction in
the spin-spin channel and $\chi^0_{\alpha \alpha'}(\vec{k},i \w_n)$
the susceptibility in the presence of the finite field $\vec{B}$
evaluated at $J=0$. Calculating these susceptibilities on the
paramagnetic side of the transition we obtain
\begin{widetext}
\begin{align} \label{effact}
S &= S_2[\vec\Phi] + S_4[\vec\Phi], \\\nonumber
S_2[\vec\Phi] &= \frac 1 \b \int \ddk{k} \sum\limits_n 
\vec{\Phi}_{\w_n,\vec{k}}^T
\begin{pmatrix}
r_\perp + |\w_n| \cos \th + k^2 & \w_n \sin \th & 0 \\
- \w_n \sin \th & r_\perp + |\w_n| \cos \th + k^2 & 0 \\
0 & 0 & r_z + |\w_n| + k^2 
\end{pmatrix}
\vec{\Phi}_{\w_{-n},-\vec{k}} \\
S_4[\vec\Phi] &= \frac {g}{\b^4} \int \ddk{k_1} \ldots \ddk{k_4} \sum\limits_{n_1 \ldots n_4} \d(\vec{k}_1 + \vec{k}_2 + \vec{k}_3 + \vec{k}_4)\d_{n_1 + n_2 + n_3 + n_4}(\vec{\Phi}_{\w_{n_1},\vec{k}_1} \cdot \vec{\Phi}_{\w_{n_2},\vec{k}_2})(\vec{\Phi}_{\w_{n_3},\vec{k}_3} \cdot \vec{\Phi}_{\w_{n_4},\vec{k}_4}).
\end{align}
\end{widetext}
Here $\b = 1/k_B T$ and $\w_n = 2 \pi n/\b$ is a Matsubara frequency
and $\vec k$ is measured again from the ordering wave vector.  The
coefficients of $\vec k^2$ and $|\w_n| \cos \th$ have been made unity
by an appropriate choice of the bare length scale $\xi_0$ and temperature/energy scale $T_0$. In
general the prefactors of the $k^2$ and $|\w|$ terms for $\Phi_z$ and
$\Phi_{x/y}$ will be different (even after rescaling), we suppress
these prefactors to keep the notations simple as they will not lead to
any qualitative changes in the results.  It is, however, essential to
keep track of the dynamics of $\Phi_{x/y}$, i.e.~of the ratio of precession and damping terms parametrized by an angle $\th$. For small $\th$ the dynamics is overdamped, while for $\th\sim \pi/2$ precession dominates. The value of $\th$ depends on details of the band-structure and the size of the magnetic field with
$\th \propto B$ for small magnetic fields.

As anticipated in (\ref{precession1}), the $x-$ and $y-$direction are
coupled for $\th>0$. The Gaussian part of the action can be
diagonalized by introducing the complex field $\Phi^\perp \equiv
(\Phi_x + \i \Phi_y)/\sqrt{2}$ as above, and we obtain
\begin{align}\label{s2}
S_2[\Phi^\perp, \Phi^z] &= \! \int \! \ddk{k} \! \frac 1\b \sum\limits_n {\Phi_{\w_n,\vec k}^\perp}^*(2 \chi_{\vec k}(\i\w_n)^{-1}) \Phi_{\w_n,\vec k}^\perp \nonumber \\
&  + \Phi^z_{\w_n,\vec k}(k^2 + r_z + |\w_n|) \Phi^z_{-\w_n,-\vec k}, 
\end{align} 
where 
\begin{equation}\label{Phipropagator}
\chi_{\vec k}(\i\w_n) \equiv (k^2 + r_\perp + |\w_n| \cos \th - \i \w_n \sin \th)^{-1}
\end{equation}
is the propagator of $\Phi^\perp$. 

As expected from the symmetry arguments given above, $r_\perp$ and
$r_z$ turn out to be different with $r_z > r_\perp$ and $r_z - r_\perp
\propto B^2$ for small $B$. As $r_{z/\perp}$ increases for increasing
fields ($r(B)\approx r(0)+c B^2$ for small $B$), an antiferromagnetic
system sufficiently close to its quantum critical point can be tuned
to the paramagnetic phase by applying a magnetic field (assuming that no first order transition is induced).

When discussing the behavior close to the quantum critical point, it
is important to note that the magnetic field enters into the
calculations both by the $B$-dependence of $r$ as well as through the
$B$-dependent angle $\th$. Close to the quantum critical point tuned
by a {\em finite} magnetic field $B_c$, $\theta(B) \approx
\theta(B_c)$ can be approximated by a constant (as checked below)
while it is obviously essential to keep track of the leading $B$
dependence of the control parameter $r(B) \propto B - B_c$.

At this point, it is worthwhile to
take a closer look at $S_2[\Phi^\perp]$ in coordinate and time space
for $\th = \pi/2$, i.e.~if Landau damping is absent as it is the case in an insulator like TlCuCl$_3$ \cite{tlcucl} or in a metal with $Q>2 k_F$ (see introduction). The Gaussian part of $S_2[\Phi^\perp]$ is minimized
for a field $\Phi^\perp$ that obeys the equation
\begin{equation}
\i \partial_t \Phi^\perp = H \Phi^\perp, \quad H = (- \nabla^2 + r_\perp). 
\end{equation}
This has the form of a Schr\"odinger equation for a particle in a
constant potential given by $V = r_\perp$. If one adds the
interactions one obtains a non-linear Schr\"odinger equation or
Gross-Pitaevskii equation which describes the physics of weakly
interacting Bosons.  In this interpretation, $r$ takes over the role
of the chemical potential.  The quantum critical point of a field
tuned insulating antiferromagnet ($\th=\pi/2$) is therefore in the
same universality class as the quantum phase transition of a dilute
gas of Bosons driven by a chemical potential. The non-magnetic phase
($r>0$) corresponds to a phase with {\em negative} chemical potential
where no Bosons are present in the $T \to 0$ limit, while the
Bose-Einstein condensed phase corresponds to the magnetically ordered
phase.

\section{Renormalization group equations and correlation length}\label{rgEq}

The physical properties of the effective action (\ref{effact}) can be
analyzed with the help of renormalization group equations.  As a first
step it is useful to perform a simple scaling analysis of
$S[\vec\Phi]$. When momenta, frequencies and fields are rescaled as
$k^\prime = k b$, $\w^\prime = \w b^z$, where $z$ is the dynamical
critical exponent, and $\Phi^\prime = \Phi b^{- \frac{d+z+2}{2}}$,
$S[\vec\Phi]$ remains invariant under scaling provided that $z=2$. The
masses $r^{\perp, z}$ and the dimensionless coupling constant $u
\equiv g\xi_0^d/T_0$ have the scaling dimensions 2 and $4-(d+z)$, respectively.
In an antiferromagnetic metal, damping as well as precession are
linear in frequency and the terms therefore behave in the same way
under scaling. In the renormalization group terminology this implies
that the precession term is an ``exactly marginal'' perturbation with
respect to Hertz fix-point ($\th=0, u=0$) which can be expected to
modify the behavior of the system at the quantum critical point.

The renormalization group equations for the parameters $T$, $r$, and
$u$ with corrections to scaling can be obtained by closely following the procedure employed by
Millis\cite{millis}: We introduce a UV-cutoff in the linked
cluster expansion of the free energy and express changes of that
cutoff in terms of changes of the parameters of the model. The RG
equations are as follows:
\begin{eqnarray}
\label{TRG1} \frac{\partial \T(b)}{\partial \log b} &= &z \, \T(b),\\
\label{rperpRG1} \frac{\partial r_\perp(b)}{\partial \log b} &= &2 r_\perp(b) + 4 u(b) \big(2 f_2^\perp (r_\perp(b), \T(b)) \big. \nonumber\\
& & + \, \big. f_2^z (r_z(b), \T(b)) \big),\\
\label{rzRG1} \frac{\partial r_z(b)}{\partial \log b} &= &2 r_z(b) + 4 u(b) \big( f_2^\perp (r_\perp(b), \T(b)) \big. \nonumber\\
& & + \, \big. 3 f_2^z (r_z(b), \T(b)) \big),\\
\label{uRG1} \frac{\partial u(b)}{\partial \log b} &= &(4 - (d+z)) u(b),
\end{eqnarray}
where $\T$ is the running temperature and the expressions for
$f_2^{\perp,z}$ as well as details of the calculation can be found in
appendix~\ref{derivation}. Since the scaling dimension for $u$ is
negative for an antiferromagnetic system in 3 spatial dimensions, we
only consider contributions up to and including first order in $u$. To
this order, the scaling law for $u$ remains unmodified, and $\th$
remains unrenormalized.  The parameter $\th$ obtains, however, finite
corrections by higher order contributions.

Equations (\ref{TRG1}) and (\ref{uRG1}) are solved trivially. As
$r_z(b)>r_\perp(b)$, $\Phi_z$ remains massive at the quantum
critical point ($r_z>0$ for $r_\perp=0$). In the following we will
concentrate on the regime $T<r_z$, where the influence of the
parallel mode $\Phi_z$ can be absorbed in a redefinition of the bare
$r_\perp$. 

Eq.~(\ref{rperpRG1}) can be solved for low temperatures in the limits
$r_\perp/T \ll 1$ and $r_\perp/T \gg 1$, corresponding to quantum
critical and (renormalized) Fermi liquid regime, respectively (see
Fig.~\ref{phaseDia}).  This provides us with an expression for the
correlation length $\xi_\perp$. We refer to appendix~\ref{derivation}
for details of the calculation. In the quantum critical regime
$\xi_\perp^{-2}$ is given by
\begin{equation} \label{rQuantCrit}
\xi_\perp^{-2}(r_\perp \ll T) = r_\perp + 16 \sqrt 2 \pi^{3/2} \, \zeta(3/2) \, u T^{3/2} \, \cos (\th/2),
\end{equation}
and in the Fermi liquid regime it has the form
\begin{equation} \label{rFermiLiquid}
\xi_\perp^{-2}(T \ll r_\perp) = r_\perp + \frac{16}{3} \pi^3 u T^2 r_\perp^{1/2} \cos \th.
\end{equation}
For all $\th<\pi/2$ one obtains the same qualitative behavior as in
the case of vanishing external magnetic field \cite{millis}. Only in
the Fermi liquid regime for $\th=\pi/2$, the $T^2$ correction is
suppressed as Landau damping is absent in this limit and our model is
characterized by an energy gap which leads to an exponential
dependence $\exp(-r_\perp/T)$ of the correlation length.

\section{Thermodynamic quantities}

In this section we calculate the specific heat $\g$, the temperature
dependence of the magnetization, 
the magnetocaloric effect $\G_B$, and the susceptibility.  The free energy can be calculated directly from RG equations
following again Ref.~[\onlinecite{millis}]. However, as the quartic coupling $u$ is irrelevant, the leading behavior in the paramagnetic phase can equivalently be extracted just from the Gaussian free energy
\begin{align}
\F & \equiv \frac{\xi_0^3}{T_0 V} (F - F(T=0)) \nonumber \\
& = - \frac{1}{4} \int \frac{d^3 k}{(2 \pi)^3} 
\int \frac{d\w}{\pi} \, \left[\coth 
\left( \frac{\b\w}{2}\right) -1 \right] \times \nonumber \\
& \quad \arctan \left(\frac{2 (r+k^2) \w \cos \th}{(r+k^2)^2 - \w^2}\right)
\label{free}
\end{align}
measured in units of $T_0 V/\xi_0^3$, and we have set $r \equiv r_\perp$.

In Eq.~(\ref{free}) and in the results shown below we 
ignore contributions from the massive,
non-critical mode $\Phi_z$ characterized by a finite mass $r_z$.  To
leading order, the corresponding (analytic) corrections to the free energy and
its derivative are just additive and can be obtained by replacing $r$
by $r_z$, by setting $\th=0$ and by dividing the result by a factor
$2$ (as there are two modes perpendicular to $B$) in all formulas for
thermodynamic quantities given below.

\subsection{Specific heat}

We first consider the specific heat coefficient $c_V/T=\g(T, r) = - \partial^2 \F/\partial T^2$.
%\begin{widetext}
%\begin{equation}
%\g_{T, r} = - \frac{\partial^2 \F}{(\partial T)^2} = \int \! \ddk{k} \int \! \frac{d x}{2 \pi} \frac{2 x}{e^x - 1} \frac{(k^2+r)^3 \left(4 (k^2+r)^2 ((k^2+r)^2+2 T^2 x^2) \cos \th + 4 T^4 x^4 \cos (3\th)\right)}{\left((k^2+r)^4 + T^4 x^4 + 2 (k^2 + r)^2 T^2 x^2 \cos(2 \th) \right)^2},
%\end{equation}
%\end{widetext}
%where $x=\b\w$. For $k \to \infty$ the integrand behaves as 
%\begin{equation}
%\frac{2 x}{e^x - 1} \frac{4 \cos\th}{k^2},
%\end{equation}
%we therefore calculate the renormalized quantity $\tilde\gamma$:
More precisely, we calculate $\tilde\gamma  \equiv \g(T, r) - \g(T=0, r=0)$:
\begin{widetext}
\begin{equation} \label{tildegamma}
\tilde\gamma = \frac 14 \int \! \ddk{k} \int \! \frac{d x}{\pi} \frac{2 x}{e^x - 1} \left[ \frac{(k^2+r)^3 \left(4 (k^2+r)^2 ((k^2+r)^2+2 T^2 x^2) \cos \th + 4 T^4 x^4 \cos (3\th)\right)}{\left((k^2+r)^4 + T^4 x^4 + 2 (k^2 + r)^2 T^2 x^2 \cos(2 \th) \right)^2} - \frac{4 \cos\th}{k^2} \right], 
\end{equation}
\end{widetext}
this differs from the physical specific heat by a $T-$independent 
(but cutoff-dependent) constant $\g_c \cos\th$. 

\begin{figure}[t]
\includegraphics*[width=\linewidth]{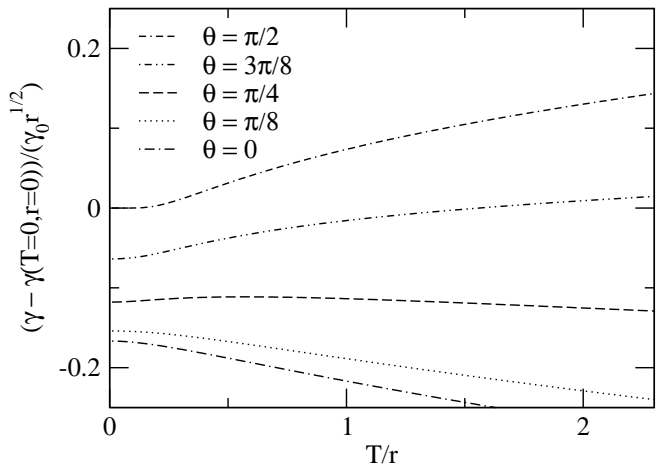}
\caption{Scaling function for the specific heat coefficient
  $\frac{1}{\sqrt{r}}\tilde\gamma$, where a non-critical contribution
  has been subtracted, $\tilde{\gamma}=\gamma(T,r)-\gamma(T=0,r=0)$.
  The function is not completely universal but depends on the
  parameter $\th$.  A crossover from $\tilde\g \propto \pm T^2$
 for $T \ll r$ to $\tilde\g \sim \pm \sqrt{T}$ for $r \ll
  T$ can be observed, where the signs depend on the value of $\th$.
  $\tilde\gamma(T)$ shows a maximum for $\pi/6 < \th <
  \pi/3$ as can be seen  more clearly in Fig.~\ref{gamma1}. \label{m}}
\end{figure}
\begin{figure}[t]
\includegraphics*[width=\linewidth]{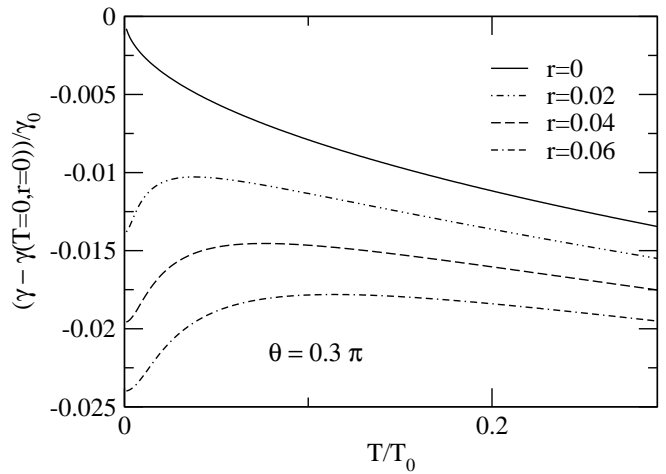}
\caption{Specific heat coefficient as a function of temperature for $\th = 0.3 \pi$ and different values of $r$. Note that the total specific heat coefficient $\gamma$ is always positive. The maximum for $r>0$ is characteristic for systems with $\pi/6 < \th <
  \pi/3$. \label{gamma1}}
\end{figure}

The integrals can be evaluated exactly in
certain limits. For $\th = \pi /2$ and low $T\ll r$ 
%\begin{equation}
%\F_{\th=\pi/2, T \to 0} = - \frac{\sqrt \pi}{(2 \pi)^3} (\sin \th)^{3/2} \, \e^{-\frac{r}{T \sin \th}} T^{5/2},
%\end{equation}
the specific heat shows thermally activated behavior
\begin{align}
\g(\th=\pi/2, T \to 0) & = \frac{\sqrt \pi}{(2 \pi)^3} \, \frac{r^2}{T^{3/2}} \exp\!\left[-\frac{r}{T}\right],
\end{align}
as can be expected from a system with a gapped spectrum. For $r \ll T$
and $T\ll r$, i.e.~in the quantum critical regime and Fermi liquid
regime, respectively, we obtain for $\th <\pi/2$
\begin{align}
\label{highT} \tilde \g(r \ll T) & = - \frac{15 \sqrt{2\pi}}{32 \, \pi^2} \zeta \left( \frac 52 \right) T^{1/2} \cos\! \left(\frac 32 \th \right), \\
\label{lowT} \tilde \g(T\ll r) & = - \frac{1}{6} r^{1/2} \cos \th - \frac{\pi^2}{60} \frac{T^2}{r^{3/2}} \cos (3 \th).
\end{align}
For $\th = 0$, this reproduces well-known results \cite{millis}
(correcting some factors of $2$), and as expected from scaling,
exponents do not change in the presence of the precession term.
However, not only the size of the prefactors but interestingly also
their sign changes when the dynamics begins to be dominated by
precession rather than damping.  In the quantum-critical regime the
$\sqrt{T}$ correction is {\em negative} for $\th<\pi/3$ and positive
for $\th>\pi/3$. Also in the Fermi liquid regime a sign change can be
observed in the $T^2/r^{3/2}$ contribution at $\th=\pi/6$.

In Fig.~\ref{m} we show the scaling function $ \frac{\tilde \g(T,
    r)}{\sqrt{r}}=f_\th(T/r)$ obtained from a numerical integration of
(\ref{tildegamma}). Due to the presence of an exactly marginal perturbation,
the scaling function is {\em not} completely universal but depends on
the parameter $\th$.  In an intermediate regime, $\pi/6 < \th <
\pi/3$, $\g(T, r)$ (and the universal scaling function $\frac{\tilde
  \g(T, r)}{\sqrt{r}}$) shows a characteristic maximum as a function
of temperature as can be read off from the asymptotical results
(\ref{highT}) and (\ref{lowT}). This maximum cannot be seen directly
at the quantum critical point ($r=0$) but for any finite $r>0$ as long
as the critical corrections to the specific heat dominate the
non-critical ones.

\subsection{Magnetization, magnetocaloric effect and Gr\"uneisen parameter}

As was argued in Ref.~[\onlinecite{gruen}], the specific heat is not
the most sensitive thermodynamic quantity close to a quantum critical
point as it tracks only variations of the free energy with respect to
temperature (vertical axis in Fig.~\ref{phaseDia}) but {\em not} with
respect to the control parameter $B$ (horizontal axis). It is
therefore interesting to study also the magnetization $M=- \partial
F/\partial B$, the susceptibility $\chi=- \partial^2 F/\partial B^2$, and the
$T$-derivative of $M$, $\partial M/\partial T=- \partial^2 F/(\partial B
\partial T)=- \partial S/\partial B$. This mixed derivative has the
advantage that -- opposite to specific heat coefficient and
susceptibility -- it vanishes in the $T \to 0$ limit due to the second
law of thermodynamics. Therefore it is not necessary to subtract any
constant non-critical contributions when measuring $\partial
M/\partial T$.

Very interesting are also ratios of the thermodynamic derivatives\cite{gruen}. For a field tuned critical point, one interesting combination is
\begin{equation}
\G_B = - \frac{(\partial M/\partial T)_B}{T \g} = - \frac 1T \frac{(\partial S / \partial B)_T}{(\partial S/\partial T)_B} = \frac 1T \left. \frac{\partial T}{\partial B} \right|_S,
\end{equation}
which describes the magnetocaloric effect, i.e.  the temperature
change in the sample after an adiabatic change of the magnetic field.

For pressure-tuned quantum phase transitions, where $\partial/\partial
B$ is replaced by $\partial/\partial p$, the quantities related to
$\partial M/\partial T$, $\chi$ and $\Gamma_B$ are the  thermal expansion, the compressibility (and therefore also to the sound
velocity), and the Gr\"uneisen parameter\cite{gruen}.

\begin{figure}
\includegraphics*[width=\linewidth]{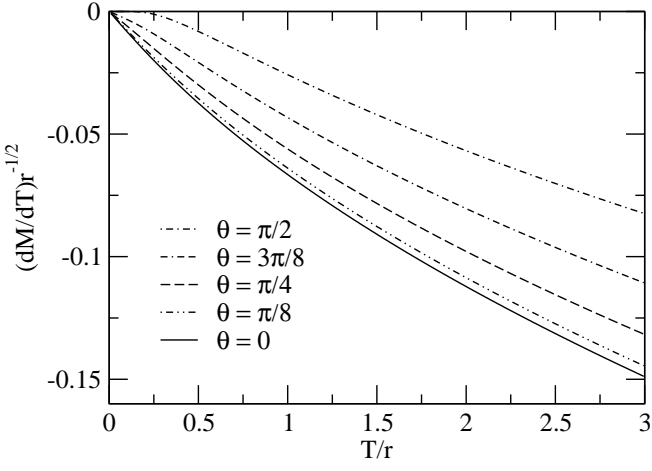}

\caption{\label{ms} Scaling function $\frac{1}{\sqrt{r}}\frac{ \partial M}{\partial T}$. While for the specific heat shown in Fig.\ref{gamma1} it was necessary to subtract a non-critical constant contribution such a background 
does not exist for $\frac{ \partial M}{\partial T}$. }
\end{figure}
\begin{figure}
\includegraphics*[width=\linewidth]{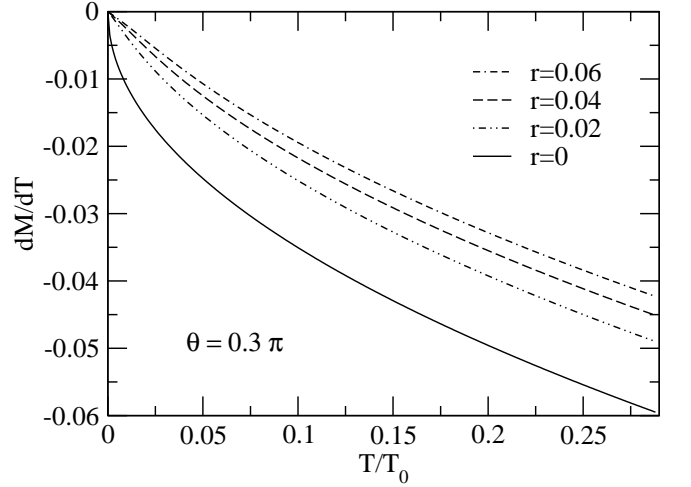}

\caption{\label{ms2} $\frac{ \partial M}{\partial T}$ as a function of temperature for $\th=0.3\pi$ and different values of $r$.
At the quantum critical point we obtain
$\frac{ \partial M}{\partial T}\sim -\sqrt{T}$ while 
$\frac{ \partial M}{\partial T} \sim -T/\sqrt{r}$ for $T \ll r$.}
\end{figure}

As both control parameter $r$ and $\th$ depend on $B$ one can expect two 
independent critical contributions to  $\partial M/\partial T$:
\begin{align}
\frac{\partial M}{\partial T} &= -\frac{\partial^2 \F}{\partial T \partial r} \frac{\partial r}{\partial B} - \frac{\partial^2 \F}{\partial T \partial \th} \frac{\partial \th}{\partial B} \nonumber \\
&= \frac 14 \int \! \ddk{k} \int \! \frac {d \w}{\pi} \left(\frac{2 e^{\w/(2 T)}}{e^{\w/T} - 1} \right)^2 \frac{\w^2}{T^2} \, \times \nonumber \\
& \bigg( \bigg.\frac{(k^2+r)^2 + \w^2) \cos \th}{((k^2+r)^4 + \w^4 + 2 (k^2 + r)^2 \w^2 \cos(2 \th)} \frac{\partial r}{\partial B} \nonumber \\ 
& + \frac{(k^2+r)^2 ((k^2+r)^2 - \w^2) \sin \th}{((k^2+r)^2 - \w^2)^2 + 4 (k^2 + r)^2 \w^2 \cos^2(\th)} \frac{\partial \th}{\partial B}\bigg. \bigg).
\end{align}
In the limits $r, T \to 0$ we obtain
\begin{align}
\frac{\partial M}{\partial T}_{r \ll T} & = - \frac{3 \sqrt {2\pi}}{16 \pi^2} \zeta \left(\frac 32 \right)\sqrt T \cos \left(\frac \th 2 \right)\frac{\partial r}{\partial B} \nonumber \\
& \quad - \frac{15 \sqrt{2\pi}}{32 \pi^2}\zeta \left(\frac 52 \right) T^{3/2} \sin \left(\frac 32 \th\right)\frac{\partial \th}{\partial B}, \\ 
\frac{\partial M}{\partial T}_{T \ll r} & = - \frac{1}{12} \frac{T}{\sqrt{r}} \cos \th \frac{\partial r}{\partial B} - \frac 16 \sqrt{r} T \sin\th \frac{\partial \th}{\partial B}.
\end{align}
As $r$ is a {\em relevant} perturbation at the quantum critical point
while $\th$ is only {\em marginal}, the contributions due to the
$B$-dependence of $\th$ are subleading and can therefore be neglected.

With $r \propto B - B_c$, we therefore find
\begin{align}
\frac{\partial M}{\partial T}_{r \ll T} & \propto
- \sqrt T  \cos \left(\frac \th 2 \right),\\
\frac{\partial M}{\partial T}_{T \ll r} & \propto - \frac{T}{\sqrt{B-B_c}} \cos \th.
\end{align} 
Neither $(\partial M/\partial T)_{r \ll T}$ nor $(\partial
M/\partial T)_{T \ll r}$ changes sign as a function of $\th$, and
indeed, $(\partial M/\partial T)_{T,r}$ is a
monotonous function of $T$ for all values of $\th$ (see Fig.~\ref{ms}).

 For $\th = \pi/2$
and at low $T$, the temperature derivative of the
magnetization also shows thermally activated behavior
\begin{equation}
\frac{\partial M}{\partial T}_{\th=\pi/2, T \to 0} = -\frac{\sqrt \pi}{(2 \pi)^3} \frac{r}{\sqrt{T}} \, \exp\left[-\frac rT \right] \frac{\partial r}{\partial B}.
\end{equation}
Finally, we evaluate the magnetocaloric effect
\begin{equation}
\G_B = - \frac{(\partial M/\partial T)_B}{T (\tilde\g + \g_c \cos\th)} ,
\end{equation}
which is given by:
\begin{align}
\Gamma_B(r \ll T) & = \frac{6 \sqrt 2 \cos(\frac\th 2) \zeta(\frac 32)}{15 \sqrt 2 \, T \cos(\frac 32 \th) \zeta(\frac 52) - 32 \sqrt{T \pi^3} \, \g_c \cos\th}, \\
\Gamma_B(T \ll r) & = \frac{1}{2(r - 6 \g_c \sqrt{r})}
\end{align}
in the limits $r, T \to 0$. Due to the non-critical contribution $\gamma_c$,
the  result for $T \to 0$ is {\em not}
fully universal\cite{gruen}.

\subsection {Susceptibility}
Since $r \sim B-B_c$ and $T$ have the same scaling exponents, one might
expect that the susceptibility $\chi=\partial M/\partial B=-\partial^2 F/\partial B^2$ and
the specific heat coefficient $\gamma= - \partial^2 F/\partial T^2$ show
very similar behavior.  This, however, turns out to be {\em not}
correct. The technical reason for this is that the susceptibility in
the quantum critical regime is a {\em singular} function of the
(dangerously irrelevant) spin-spin interaction $u$. Practically, this
implies that a measurement of the susceptibility is complementary to
other thermodynamic measurements as it is highly sensitive to a quantity
which can otherwise be determined only by neutron scattering measurements
of the correlation length.

The susceptibility $\chi=-\partial^2 F/\partial B^2$ gets contributions
both from the $B$-field dependence of $\theta$ and of $r$. We only
consider the leading corrections due to $r$ (see discussion above) and
evaluate the quantity $\tilde{\chi}(r, T)=\chi(r, T) - \chi(r=0, T=0)$ with
\begin{align}
\tilde{\chi}&=\int \! \! \ddk{k} \int \! \! \frac{d \w}{\pi} \Big[ \Big. n_B (\w) {\rm Im} \frac{1}{(k^2+r+\i \w \cos\th - \w \sin\th)^2} \nonumber\\
& \quad + \Theta(-\w) {\rm Im} \frac{1}{(k^2+\i \w \cos\th - \w \sin\th)^2} \Big. \Big].
\end{align}

Most interesting is the quantum critical regime, where the leading
correction to the susceptibility takes the form
\begin{equation} \label{chiQuantCrit}
\tilde{\chi} (r \ll T) \approx \left(\frac{1}{8 \pi} \frac{T}{\sqrt{r}} - \frac{\sqrt 2}{4 \pi^2} \sqrt T \cos\left(\frac\th 2\right)\right) 
\left(\frac{\partial r}{\partial B}\right)^2.
\end{equation}

Note that this expression formally {\em diverges} for $r \to 0$. This
implies that we have to take into account the interaction effects
discussed in Section~\ref{rgEq} and we have to replace the control parameter
 $r$ by $\xi_\perp^{-2}(T)\sim r + u T^{3/2}$ given by Eq.~(\ref{rQuantCrit}). One therefore finds 
\begin{equation} \label{chiQuantCrit3}
\tilde{\chi} \propto \frac{T}{\sqrt{r}} \qquad \text{for }\  r< T<(r/u)^{2/3},
\end{equation}
but 
\begin{equation} \label{chiQuantCrit2}
\tilde{\chi} \propto \frac{T^{1/4}}{\sqrt{u}} \qquad \text{for } \ 
T>(r/u)^{2/3}.
\end{equation}
Scaling is violated, i.e.~the susceptibility is no longer of the form
$\tilde\chi(r,T) = f_\th (T/r)/\sqrt{r}$, as the dangerously irrelevant
coupling $u$ determines $\tilde\chi$ in this regime. It is interesting to
trace back the origin of the $1/\sqrt{r}$ contribution in
(\ref{chiQuantCrit}). It arises from the $\w_n=0$ mode of the Gaussian
theory (\ref{s2}) which leads to a contribution of the form $T \sum_k
\left(\frac{1}{k^2+r}\right)^2$ to the susceptibility. Note that the
static $\w=0$ contribution does not depend on the dynamics, i.e.~it
does not depend on $\theta$. However, the correlation length $\xi_\perp(T)$ 
given in Eq.~(\ref{rQuantCrit}) does depend smoothly on $\theta$ which leads to a slight $\theta$-dependence of $\tilde\chi$.

% This is illustrated in Fig. \ref{},
% where the temperature dependence of $\chi(r, T)$ is plotted for fixed
% $r$ and different values of $u$. 

In the Fermi liquid regime, the susceptibility is given by
\begin{equation}
\tilde{\chi} (T \ll r) = \left(- \frac {1}{4 \pi^2} \sqrt r \cos\th + \frac {1}{48} \frac{T^2}{r^{3/2}} \cos\th\right) \left(\frac{\partial r}{\partial B}\right)^2,
\end{equation}
while for $\th = \pi/2$ and low $T \ll r$ it shows thermally activated behavior
\begin{equation}
\chi (\th = \pi/2, T \to 0) = \left(\frac{\sqrt \pi}{(2 \pi)^3} \sqrt T \exp\left[- \frac rT \right]\right) \left(\frac{\partial r}{\partial B}\right)^2.
\end{equation}

\begin{figure}
\includegraphics*[width=\linewidth]{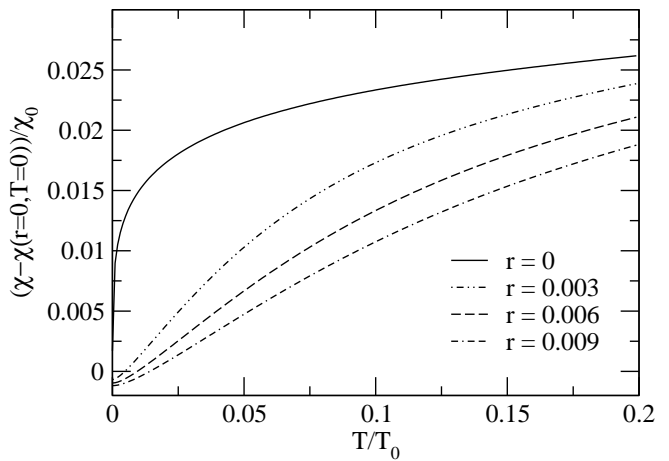}
\caption{Susceptibility as a function of temperature for $u = 0.2$, $\th = 0.01$, and different values of $r$. Curves for other values of $\theta$ look essentially identical and are not shown. Note that the susceptibility is {\em much more} sensitive to small deviations from the quantum critical point than the specific heat coefficient or $\partial M/\partial T$ (c.f.~Fig.~\ref{gamma1}, \ref{ms2}, where much larger values for $r$ have been used). The $T^{1/4}$ cusp at the quantum critical point [Eq.~\ref{chiQuantCrit2}] is rapidly washed out by tiny deviations from the critical magnetic field and replaced by the linear dependence of Eq.~(\ref{chiQuantCrit3}).
\label{suscSmallAnglePlot}}
\end{figure}

The rapid crossovers between the $T^2$, $T$ and $T^{1/4}$ regimes are shown in 
Fig.~\ref{suscSmallAnglePlot}.

%For certain values of the coupling constant $u$ and low temperatures it can happen that the the next-to leading term in $\chi(r, T)$ dominates the leading term for certain temperatures, i.e.~$\chi(r, T)$ shows a $\sqrt T$-dependence in that temperature region. Since the $\sqrt T$-contribution is negative (\ref{chiQuantCrit}), this leads to a marked difference in the shape of the $\chi-T$-curve, as can be seen in Fig.~\ref{suscBigU}: When the system is tuned to the quantum critical point, the susceptibility exhibits local extrema as a function of temperature. 

%\begin{figure}
%\includegraphics*[scale=0.6]{suscBigU.eps}
%\caption{Susceptibility as a function of temperature for $u = 0.8$, $\th = 0.1 \pi$, and different values of $r$.\label{suscBigU}}
%\end{figure}

\section{Scattering rate}

Due to energy and momentum conservation, the scattering of electrons
from spin-fluctuations is most efficient close to ``hot lines'' on the
Fermi surface, where $E_{\vec{k}}=E_{\vec{k}\pm \vec{Q}}=0$, where
$\vec{Q}$ is the ordering vector of the antiferromagnet.  In order to
determine how spin precession modifies the results we calculate the
scattering rate as a function of $\th$. We will neglect all orbital
effects of the magnetic field and will not try to calculate the
conductivity and the Hall effect. For an extensive discussion of
orbital effects and magnetotransport in nearly antiferromagnetic
metals see Ref.~[\onlinecite{rosch:2000}] which does, however, not
consider effects of spin precession.

In second order perturbation theory the lifetime of the spin-up
electron scattering from fluctuations of $\Phi_\perp$ at $T=0$ is
given by \cite{hlubina}
\begin{equation}
\frac {1}{\tau^{\uparrow}_{\vec k}} = 2 g_s^2 \sum\limits_{\vec {k^\prime}} \int_0^{\epsilon_{\vec{k}}} \! d\w \, {\rm Im} \chi_{\vec k - \vec{k^\prime}}(\w) \d[\w - (E^+_{\vec k} - E^-_{\vec{k^\prime}})],
\end{equation}
where $g_s$ is a coupling constant,  $E^{+/-}_{\vec k}$  and $v_F^{+/-}$
the energy and velocity of spin up/down electrons and $\chi$ is the spin fluctuation spectrum of Eq.~(\ref{Phipropagator}),
\begin{equation}
\chi_{\vec q}(\w) = \frac{1}{\w_{\vec q} + r + \i \w \cos \th - \w \sin \th},
\end{equation}
with $\w_{\vec q} = (\vec{q}\pm \vec{Q})^2/q_0^2$. We split the
momentum integration in an integral over the
Fermi surface and an energy integration $\int \! d^3\vec{k^\prime} = \int \! \int \! d\vec{k^\prime}/v_F^- \int d E^-_{k^\prime}$ and integrate 
 first over $E^-_{k^\prime}$, then over $\w$ to obtain
\begin{widetext}
\begin{align} \label{tauup}
\frac {1}{\tau^{\uparrow}_{\vec k}} &\approx \frac{g_s^2}{v_F^- (2 \pi)^3} \int \! \int \! d\vec{k^\prime} \, \left( \cos \th \ln \left[\frac{(\w_{\vec k - \vec{k^\prime}}+r)^2 + 2 E^+_{\vec k} \sin\th + (E^+_{\vec k})^2} {(\w_{\vec k - \vec{k^\prime}}+r)^2} \right] + \sin \th \arctan \left[\frac{- E^+_{\vec k} \cos \th }{\w_{\vec k - \vec{k^\prime}}+r + E^+_{\vec k} \sin \th} \right] \right) \\
&\approx \frac{g_s^2 q_0^2}{v_F^- (2 \pi)^2} E^+_{\vec k} {\rm min} \left\{\frac{E^+_{\vec k}}{2 \d_k^2} \cos \th, \frac \pi 2 - \th\right\}, 
\end{align}
\end{widetext}
where $\d_k = r + (\d\vec{k}/q_0)^2$ and $\d\vec{k}$ is the distance of $\vec k + \vec Q$ from the Fermi
surface or, approximately, the distance of $\vec k$ from hot lines on
the Fermi surface. Analogously we obtain for spin-down electrons
\begin{equation}
\frac {1}{\tau^{\downarrow}_{\vec k}} \approx \frac{g_s^2 q_0^2}{v_F^+ (2 \pi)^2} E^-_{\vec k} {\rm min} \left\{\frac{E^-_{\vec k}}{2 \d_k^2} \cos \th, \frac \pi 2 - \th\right\}, 
\end{equation}
where the indices $+$ and $-$ have been exchanged w.r.t.
(\ref{tauup}).  The scattering rate is strongly dependent on the
distance from the hot lines: at the quantum critical point and for
$\d\vec{k}/q_0 \approx 0$ the scattering rate is linear in the
quasiparticle energy. Far away from the hot lines and the quantum
critical point the usual scattering rate $1/\tau^{\uparrow,
  \downarrow}_{\vec k} \propto {E^{+,-}_{\vec k}}^2$ is recovered
\cite{rosch:2000}. The main result of this section is that the
spin-precession term does not lead to a qualitative change in the
scattering rate.

\section{Discussion}

In this paper, we have discussed the field-induced quantum phase
transition of a clean, three-dimensional antiferromagnetic metal,
restricting our attention to the non-magnetic side of the phase
diagram.  The main question was how the interplay of precession of the
spins in the presence of a finite magnetic field and Landau damping
modifies the quantum critical behavior.

One main qualitative result of our analysis is that the critical
behavior is {\em not} completely universal as it depends on a continuous
variable $\th$ which parametrizes the ratio of precession and damping
terms in the effective action. While critical exponents do not depend
on $\th$, this parameter strongly changes the scaling functions and
even the sign of leading corrections e.g. to the specific heat.

A requirement for the validity of our analysis is that the two modes
$\Phi_{x,y}$ perpendicular to the magnetic field are characterized by
the same mass. In the presence of sizable spin-orbit couplings, this
will only be the case, if the crystal has a sufficiently high symmetry
and if furthermore the external magnetic field is applied along a
symmetry direction of a crystal.

Presently, we are not aware of any experiments which show for example
the maximum in the $T$ dependence of the specific heat coefficient
which we predict for $\pi/6<\th<\pi/3$. Note that in systems like
CeCu$_{5.2}$Ag$_{0.8}$ \cite{heuser98} or CeCu$_{5.8}$Au$_{0.2}$
\cite{loehneysen01} strong anisotropies prohibit the precession of the
spin, i.e.~$\theta=0$. Under what conditions can large values of $\th$
be expected? Obviously, large uniform magnetizations are required.
In heavy Fermion systems with Kondo temperatures of the
order of a few Kelvin, one can introduce strong magnetic polarizations
with moderate external fields and it should therefore be possible to
induce sizable values of $\th$.  A different class of systems which
might be of interest in this context are ferrimagnetic materials. If
it is possible to suppress only the staggered component of the
magnetization in such systems either by external fields, pressure or
doping, the critical theory within the Hertz approach will be
characterized by a finite (and again sizable) $\th$.

As long as the ordering vector $\vec{Q}$ of the antiferromagnet can
connect the spin-up and spin-down Fermi surfaces
($Q<k_F^++k_F^-$ or, more precisely, $E^+_{\vec{k}}=E^-_{\vec{k}\pm \vec{Q}}=0$ for a line of momenta
$\vec{k}$), Landau damping is present and $\th < \pi/2$. In contrast,
one finds $\th=\pi/2$ in all systems where no such connection exists
($Q>k_F^++k_F^-$).
However, the transition from $\th < \pi/2$ to $\th=\pi/2$ is {\em not}
expected to be smooth, as the interactions of the spin-fluctuations diverge and become relevant\cite{millis,2kf} at the point where  $Q=k_F^++k_F^-$.

\begin{figure}
\includegraphics*[width=0.9 \linewidth]{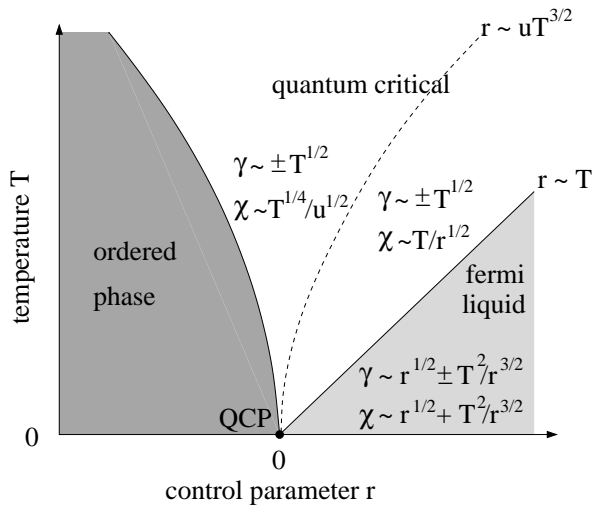}

\caption{\label{overview} Qualitative behavior of specific heat coefficient $\gamma$ and magnetic susceptibility $\chi$ in various regimes of the phase diagram. }
\end{figure}

According to our analysis, the susceptibility $\partial M/\partial B$
is a particularly interesting experimental quantity to study close to
a field-driven quantum critical point. First of all, it is expected to
be much more sensitive to small deviations from criticality compared to
other thermodynamic quantities (see Fig.~\ref{suscSmallAnglePlot}).
Second, it allows to measure the correlation length, a quantity which
cannot be extracted from other thermodynamic quantities, as for $B=B_c$
we obtain from (\ref{chiQuantCrit})
\begin{equation}
\frac{\chi(T)-\chi(T=0)}{T} \propto \xi(T).
\end{equation}
Third, it strongly violates the $T/(B-B_c)$ scaling. This deviation
from scaling for $\chi$ can be used to show that the relevant critical
theory is above its upper critical dimension, a central question for the interpretation of quantum criticality in systems like CeCu$_{6-x}$Au$_x$ or YbRh$_2$Si$_2$ \cite{loehneysen01,steglich}. All these three statements 
are actually independent of the value of $\theta$: they apply equally for a dynamics which is overdamped, $\theta \ll 1$, or for a BEC system like TlCuCl$_3$ with $\theta=\pi/2$. Note that in pressure-tuned quantum critical points the 
compressibility $\kappa$ (and therefore the sound velocity\cite{luethi})
plays the same role as $\chi$ for field tuned quantum phase transitions.
An overview of the qualitative $T$ dependence of the specific heat and susceptibility is shown in Fig.~\ref{overview}.

Field tuned quantum-phase transitions in metals allow to study quantum
critical behavior with a tuning parameter which can easily be
controlled and with a conjugate field -- the uniform magnetization --
which can directly be measured. They are therefore especially well
suited to answer some of the central questions in the field of quantum
critical metals, for example, whether or not such systems can be
described in terms of simple spin-fluctuation theories as have been
used in this paper.

\acknowledgements

We thank T. Lorenz, B. Luethi, S. Paschen, A. Schofield, N. Shah, Q. Si, and
M. Vojta for useful discussions and the Emmy-Noether program and SFB
608 of the DFG for financial support.

\appendix
\section{Cubic terms in the effective action}\label{cubic}
In this appendix we briefly discuss whether cubic terms $\Phi^3$
are present in the low-energy effective Lagrangian. As $\Phi$ carries the
momentum $\vec{Q}$, the presence of such terms is
forbidden by momentum conservation in most systems with the exception
of magnetic structures (e.g. BCC lattices) where the sum of three
ordering vectors adds up to 0. If such a system has Ising symmetry,
then a cubic term does exist and the magnetic field driven transition
will be first order. However, for $xy$ symmetry perpendicular to the
magnetic field (the case mostly discussed in this paper), a
rotationally invariant cubic term of the form $\Phi_\perp^3$ does {\em
  not} exist. While  terms like $B \Phi_z |\Phi_\perp|^2$ {\em are}
allowed by symmetry, they lead effectively only to a renormalization of the 
$|\Phi_\perp|^4$ interaction as $\Phi_z$ remains massive. We therefore neglect such terms.

\section{Derivation of RG-equations}\label{derivation}

Following Millis' treatment\cite{millis,markusDA}, we perform
the renormalization group analysis on the free energy after having
converted all Matsubara sums to integrals. Although we restrict our
calculations to systems in three spatial dimensions and with a dynamical
critical exponent of $z=2$, we nonetheless keep the variables $d$ and
$z$ in the calculation in order to make the origin of certain factors more
transparent.

\begin{figure}
\includegraphics*{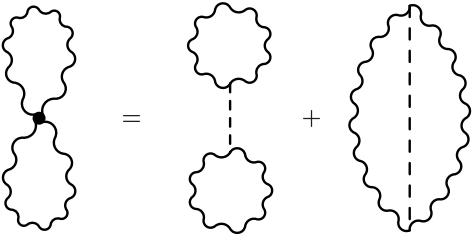}
\caption{Diagram contributing to the free energy in ${\mathcal O}(u)$. The different contractions of the internal indices of the fields involved have been made explicit on the right hand side, where the dashed line represents the quartic interaction $u$.\label{freeenergydiag}}
\end{figure}

The free energy can be obtained via a linked cluster expansion in the
coupling constant $u$. The scaling dimension of $u$ is $4- (d+z)$, which is negative for an
antiferromagnetic system in 3 spatial dimensions. To first order in $u$, only the diagram in Fig.~\ref{freeenergydiag} contributes to the free energy. 
Up to first order in $u$, the free energy $\F$ is therefore given by
\begin{equation} \label{freeenexp1}
\F = \F_G + u [(I_{\chi} + I_{\chi^*} + I_{\chi^z})^2 + 2(I_{\chi}^2 + I_{\chi^*}^2 + I_{\chi^z}^2)],
\end{equation}
where 
\begin{align} \label{fg}
\F_G &= - \frac{1}{2} \int \limits_0^\Lambda \frac{d^3 k}{(2 \pi)^3} \int \limits_0^\Gamma \frac{d\w}{\pi} \, \coth \left( \frac{\b\w}{2}\right) \times \nonumber \\
& \qquad \arctan \left(\frac{2 (r+k^2) \w \cos \th}{(r+k^2)^2 - \w^2}\right)
\end{align}
is the Gaussian free energy measured in units of $T_0 V/\xi_0^3$ with the cutoffs $\Lambda$ and $\Gamma$, $I_{\chi}$ is given by
\begin{align} \label{idiag}
I_{\chi} &\equiv \int \! \ddk{k} \frac 1\b \sum\limits_n \chi_{\vec k}(i\w_n) \nonumber \\
&= \int \limits_0^\Lambda \ddk{k} \int \limits_0^\Gamma \frac{d\w}{\pi} \coth \left(\frac{\b \w}{2}\right) \times \nonumber \\
& \quad \frac{\w \cos \th}{(r+k^2)^2-2(r+k^2)\w \sin\th + \w^2} ,
\end{align}
and $I_{\chi^*}$, $I_{\chi^z}$ are defined analogously. 

As a next step, we separate out of the momentum and frequency
integrals in the expressions on the right hand side of (\ref{freeenexp1}) the regions given by $\{\Lambda \geq k \geq \Lambda/b,\Gamma \geq
\w \geq 0\}$ and $\{\Lambda \geq k \geq 0,\Gamma \geq \w \geq
\Gamma/b^2\}$. Using that 
\begin{equation}
I_{\chi} + I_{\chi^*} = 2 \frac{\partial \F_G}{\partial r_\perp}, \ I_{\chi^z} =  2 \frac{\partial \F_G}{\partial r_z},
\end{equation}
the change in $\F$ upon such a variation of the cutoff can be
expressed as a change of $r_\perp$ and $r_z$, and this leads to the
equations
\begin{eqnarray}
\label{rperpRG2} \frac{\partial r_\perp(b)}{\partial \log b} &= &2 r_\perp(b) + 4 u(b) \big( 2 f_2^\perp (r_\perp(b), \T(b)) \big. \nonumber\\
& & + \, \big. f_2^z (r_z(b), \T(b)) \big),\\
\frac{\partial r_z(b)}{\partial \log b} &= &2 r_z(b) + 4 u(b) \big( f_2^\perp (r_\perp(b), \T(b)) \big. \nonumber\\
& & + \, \big. 3 f_2^z (r_z(b), \T(b)) \big)
\end{eqnarray}
for the running masses $r_\perp(b)$, $r_z(b)$, where $f_2^\perp$ and $f_2^z$ are given by
\begin{align}
f_2^\perp(r_\perp, \T) &= K_3 \Lambda^3 \int\limits_0^\Gamma \frac{d\w}{\pi} \coth\left(\frac{\b \w}{2} \right) \times \nonumber \\
& \frac{2 \w ((\Lambda^2 + r_\perp)^2 + \w^2)\cos\th}{((\Lambda^2+r_\perp)^2 + \w^2)^2 - 4 (\Lambda^2+r_\perp)^2 \w^2 \sin^2 \th} \nonumber \\
& + \frac{z \Gamma}{\pi} \int \limits_0^\Lambda \ddk{k} \coth\left(\frac{\b \Gamma}{2} \right) \times \nonumber \\
& \frac{2 \Gamma ((k^2 + r_\perp)^2 + \Gamma^2)\cos\th}{((k^2+r_\perp)^2 + \Gamma^2)^2 - 4 (k^2+r_\perp)^2 \Gamma^2 \sin^2 \th}, \nonumber\\
f_2^z(r_z, \T) &= K_3 \Lambda^3 \int\limits_0^\Gamma \frac{d\w}{\pi} \coth\left(\frac{\b \w}{2} \right) \frac{\w}{(r_z+\Lambda^2)^2 + \w^2}\nonumber \\
& + \frac{z \Gamma}{\pi} \int \limits_0^\Lambda \ddk{k} \coth\left(\frac{\b \Gamma}{2} \right) \frac{\Gamma}{(r_z+k^2)^2 + \Gamma^2}.
\end{align}
In the following we assume that the system is close to the quantum
critical point at temperatures much smaller than $r_z$. In this case,
$f_2^z(r_z, \T)$ can be set to zero, and the renormalization group flow
of $r_\perp$ is determined by $f_2^\perp (r_\perp, \T)$ only. There are
two contributions to $f_2^\perp$, one from the renormalization due to
the separated momentum shell, where momentum is set on shell $k =
\Lambda$, and one from the renormalization due to the frequency shell
with $\w = \Gamma$.  For subsequent calculations we note that
\begin{align}
f_2^\perp(r_\perp, \T ) - f_2^\perp(r_\perp, 0) &= K_3 \Lambda^3 \int\limits_0^\Gamma \frac{d\w}{\pi} \left[\coth\left(\frac{\b \w}{2}\right) - 1\right] \times \nonumber \\
& \hspace{-3.5cm}\frac{2 \w ((\Lambda^2 + r_\perp)^2 + \w^2)\cos\th}{((\Lambda^2+r_\perp)^2 + \w^2)^2 - 4 (\Lambda^2+r_\perp)^2 \w^2 \sin^2 \th} + \O (e^{-\Gamma/\T}), 
\end{align}
in other words the contribution of the frequency shell renormalizes
zero temperature properties only and is exponentially suppressed at
finite temperatures.
 
In order to obtain an expression for the correlation length, we first
substitute $r_\perp(b) = R_\perp(b) b^2$ to eliminate the naive scaling and then
formally integrate equation (\ref{rperpRG2}):
\begin{equation}
R_\perp(b) = r_0^\perp + 8 \int\limits_0^{\ln b} \! dx e^{-2 x} u(e^x) f_2^\perp(R_\perp(e^x) e^{2x}, T e^{zx}).
\end{equation}
We then perform an expansion in temperature
\begin{equation} \label{Texpansion}
R_\perp(b) \sim \Delta_\perp(b) + R_T^\perp(b) + \delta R_\perp(b),
\end{equation}
where three terms contribute.

The first term $\Delta_\perp(b)$ is the running mass at zero temperature,
\begin{equation}
\Delta_\perp(b) = r_0^\perp + 8 \int\limits_0^{\ln b} \! dx \, e^{-2 x} u(e^x) f_2^\perp(\Delta_\perp(e^x) e^{2x}, 0);
\end{equation}
the integrand can now be expanded in $\Delta_\perp$ which leads to the following expression
\begin{eqnarray}
\Delta_\perp(b) & \sim & r_0^\perp + 8 f_2^\perp(0, 0) \int\limits_0^{\ln b} \! dx \, e^{-2 x} u(e^x) \stackrel{b \rightarrow \infty}{\rightarrow} r_0^\perp - r_c^\perp \nonumber \\
&\equiv & r_\perp ,
\end{eqnarray}
i.e. this defines the parameter which characterizes the distance from
the critical point. 

The other two terms in (\ref{Texpansion}) are of first order in
temperature: One contribution is due to an explicit dependence of
$f_2^\perp$ on the running temperature,
\begin{eqnarray}
R_T^\perp(b) &=& 8 \int\limits_0^{\ln b} \! dx e^{-2 x} u(e^x) \big(f_2^\perp(R_\perp(e^x) e^{2x}, T e^{z x}) \big. \nonumber\\
&& \big. - f_2^\perp(R_\perp(e^x) e^{2x}, 0)\big),
\end{eqnarray} 
and $\delta R_\perp(b)$ originates from the temperature dependence of the running mass,
\begin{eqnarray}
\delta R_\perp(b) &=& 8 \int\limits_0^{\ln b} \! dx e^{-2 x} u(e^x) \big(f_2^\perp(R_\perp(e^x) e^{2x}, 0) \big. \nonumber\\
&& \big. - f_2^\perp(\Delta_\perp(e^x) e^{2x}, 0) \big).
\end{eqnarray}
This term is of the order of $u^2$ and will be neglected from now on.

The inverse square of the correlation length $\xi_\perp$ is given by
\begin{widetext}
\begin{eqnarray}
\xi_\perp^{-2} &=& \lim_{b \rightarrow \infty} \{\Delta(b) + R_T^\perp(b)\} = r_\perp + \lim_{b \rightarrow \infty} 8 \int\limits_0^{\ln b} \! dx \, e^{-2 x} u(e^x) K_3 \Lambda^3 \times \nonumber \\
&&  \int\limits_0^\Gamma \frac{d\w}{\pi} 
\left[\coth\left(\frac{\w}{2 T \e^{z x}}\right) - 1\right]
\frac{2 \w ((\Lambda^2 + R_\perp(e^x) e^{2x})^2 + \w^2)\cos\th}{((\Lambda^2+R_\perp(e^x) e^{2x})^2 + \w^2)^2 - 4 (\Lambda^2+R_\perp(e^x) e^{2x})^2 \w^2 \sin^2 \th} \\
& = &  \label{uglyXi}
r_\perp + 16 \Lambda^{d+z-2} K_d T^{2/z} \int_{\ln(\frac{T^{1/z}}{\Lambda})}^\infty d x \, u(\e^x \Lambda T^{-1/z}) \e^{(z-2) x} \int_0^\infty \frac{d v}{\pi} (\coth v - 1) \\
&&\frac{4 \Lambda^z \e^{z x} v \big((\Lambda^2 + R_\perp(e^x \lambda T^{1/z}) e^{2 x} \Lambda^2 T^{-2/z})^2 + 4 \Lambda ^{2 z} \e^{2 z x} v^2 \big) \cos \th}{\big((\Lambda^2 + R_\perp(e^x \lambda T^{1/z}) e^{2 x} \Lambda^2 T^{-2/z})^2 + 4 \Lambda ^{2 z} \e^{2 z x} v^2 \big)^2 - 16(\Lambda^2 + R_\perp(e^x \lambda T^{1/z}) e^{2 x} \Lambda^2 T^{-2/z})^2 v^2 \Lambda^{2 x} \e^{2 z x}\sin^2 \th}, \nonumber
\end{eqnarray}
\end{widetext}
where the transformations 
$\e^{x^\prime} = \e^x \Lambda^{-1} T^{1/z}$ and 
$ v = \w/2 T \e^{z x}$  
% = \frac{\w}{2 \Lambda^z \e^{z x^\prime}}
have been introduced, and $u(\e^x \Lambda T^{-1/z})$ = $u_0 (\e^x
\Lambda T^{-1/z})^{4-(d+z)}$. Expression (\ref{uglyXi}) for the
correlation length can now be evaluated in the quantum critical and
Fermi liquid regime.
 
In the quantum critical regime we can neglect the dependence of the
integrand of (\ref{uglyXi}) on $R_\perp$ and extend the lower limit of
the $x$-integral to $-\infty$.  Using the following integral
\begin{equation}
\int_0^\infty d\xi \, \frac{(2 \xi)^n}{\sinh^2 \xi} = 2 n \Gamma(n) \zeta(n), \ n = 0, 1, 2, \ldots
\end{equation}
we obtain 
\begin{eqnarray}
\xi_\perp^{-2}&=& r_\perp + 16 \frac{K_d}{z \cos \left(\frac{d-2}{2 z} \pi \right)} \Gamma\left(1 + \frac{d-2}{z}\right) \zeta\left(1 + \frac{d-2}{z}\right)\nonumber \\
&& \times u \, T^{\frac{d+z-2}{z}} \cos \left(\frac{d-2}{z} \th \right).
\end{eqnarray}
%Note that the RG calculation with approximations as those above is
%identical to the result that we would have obtained from a calculation
%of the graph .

In the Fermi liquid regime and for low temperatures, we can replace
the running mass $R_\perp$ in (\ref{uglyXi}) by the control parameter
$r_\perp$. It is convenient at this point to introduce yet another
variable transformation of the form $\e^{2 x^\prime} = r_\perp
T^{-2/z} \e^{2 x}$. To lowest order we can then neglect the term $T
r_\perp^{-z/2}$ in the integrand. Furthermore, we can extend the lower limit
of the $x$-integral to $- \infty$, thereby inducing an error of order
$\O(r_\perp^{1/2}/\Lambda)^{2-d+z}$, and obtain
\begin{equation}
\xi_\perp^{-2} = r_\perp + 16 \frac{\pi^2}{12} \frac{d-z}{\sin\left( \frac{d-z}{2} \pi \right)} K_d u T^2 r_\perp^{\frac{d-z-2}{2}} \cos \th
\end{equation}
in the Fermi liquid regime.


\begin{thebibliography}{99}

\bibitem{heuser98} K. Heuser, E.-W. Scheidt, T. Schreiner, and G.R. Stewart, Phys. Rev. B {\bf 57}, R4 198; K. Heuser, E.-W. Scheidt, T. Schreiner, and G.R. Stewart, Phys. Rev. B {\bf 58}, R15 959.

\bibitem{loehneysen01} H. von L\"ohneysen, C. Pfleiderer, T. Pietrus, O. Stockert, and B. Will, Phys. Rev. B {\bf 63}, 134411 (2001).

\bibitem{bauer00} E. Bauer, A. Galatanu, L. Naber, M. Galli, F. Marabelli, C. Seuring, K. Heuser, E.-W. Scheidt, T. Schreiner, and G. R. Stewart, Physica B {\bf 281-282}, 319 (2000).

\bibitem{steglich} J. Custers, P. Gegenwart, H. Wilhelm, K. Neumaier, Y. Tokiwa, O. Trovarelli, C. Geibel, F. Steglich, C. P\'epin, P. Coleman,
Nature {\bf 424}, 524 (2003); 
P. Gegenwart, J. Custers, C. Geibel, K. Neumaier, T. Tayama, K. Tenya, O. Trovarelli, and F. Steglich, Phys. Rev. Lett. {\bf 89}, 056402 (2002).
% Magnetic-Field Induced Quantum Critical Point in YbRh2Si2

\bibitem{CeIrIn} J. S. Kim, N. O. Moreno, J. L. Sarrao, J. D. Thompson, and G. R. Stewart, Phys. Rev. B 69, 024402 (2004).

\bibitem{sarrao} J. L. Sarrao {\it et al.}, unpublished.

\bibitem{paglione} J. Paglione, M.A. Tanatar, D.G. Hawthorn, E. Boaknin, F. Ronning, R.W. Hill, M. Sutherland, Louis Taillefer, C. Petrovic, and P.C. Canfield, cond-mat/0405157.

\bibitem{tlcucl} A. Oosawa, M. Ishii and H. Tanaka, J. Phys.: Condens.
  Matter {\bf 11}, 265 (1999); C. R\"uegg {\it et al.},
% N. Cavadini, A. Furrer, H.-U. G\"udel, K. Kr\"amer,
%H. Mutka, A. Wildes, K. Habicht und P. Vorderwisch},
Nature {\bf 423}, 62 (2003).

\bibitem{srcubo}
%Exact Dimer Ground State and Quantized Magnetization Plateaus
% in the Two-Dimensional Spin System SrCu2(BO3) 2
H. Kageyama {\it et al.},
%1,2*, K. Yoshimura1,3?, R. Stern3, N. V. Mushnikov2, K. Onizuka2, M. Kato1, K. Kosuge1, C. P. Slichter3, T. Goto2, and Y. Ueda2 
Phys. Rev. Lett. {\bf 82}, 3168 (1999).

\bibitem{BaCuSi2O6}
%Magnetic-Field-Induced Condensation of Triplons in Han Purple Pigment BaCuSi2O6
M. Jaime {\it et al.},
%1 V. F. Correa,1 N. Harrison,1 C. D. Batista,2 N. Kawashima,3 Y. Kazuma,3 G. A. Jorge,1,4 R. Stern,5 I. Heinmaa,5 S. A. Zvyagin,6 Y. Sasago,7 and K. Uchinokura
 Phys. Rev. Lett. {\bf 93}, 087203 (2004).

% \bibitem{stewart} G. R. Stewart, Rev. Mod. Phys. {\bf 73}, 797 (2001).

\bibitem{metamagneticQPT}

S. A. Grigera, R. S. Perry, A. J. Schofield, M. Chiao, S. R. Julian, G. G. Lonzarich, S. I. Ikeda, Y. Maeno, A. J. Millis, A. P. Mackenzie, 
Science, {\bf 294}, 329 (2001);
A. J. Millis, A. J. Schofield, G. G. Lonzarich, S. A. Grigera,
Phys. Rev. Lett. {\bf 88}, 217204 (2002);
R. S. Perry, K. Kitagawa, S. A. Grigera, R. A. Borzi, A. P. Mackenzie, K. Ishida, and Y. Maeno, Phys. Rev. Lett. {\bf 92}, 166602 (2004).

\bibitem{moriyaBuch} T. Moriya, {\it Spin Fluctuations in Itinerant Electron Magnetism}, Springer Verlag, Berlin (1985).

\bibitem{hertz} J.A. Hertz, Phys. Rev. B {\bf 14}, 1165 (1976).

\bibitem{millis} A.J. Millis, Phys. Rev. B {\bf 48}, 7183 (1993).

\bibitem{CeCuAu}
H. v. L\"ohneysen, T. Pietrus, G. Portisch, H.G. Schlager, A. Schr\"oder, M. Sieck, and T. Trappmann, Phys. Rev. Lett. {\bf 72}, 3262 (1994); A. Rosch, A. Schr\"oder, O. Stockert, and H. v. L\"ohneysen, Phys. Rev. Lett. {\bf 79}, 159 (1997); O. Stockert, H. v. L\"ohneysen, A. Rosch, N. Pyka, and M. Loewenhaupt, {\it ibid.} {\bf 80}, 5627 (1998); A. Schr\"oder, G. Aeppli, R. Coldea, M. Adams, O. Stockert, H. v. L\"ohneysen, E. Bucher, R. Ramazasavili, and P. Coleman, Nature {\bf 407}, 351 (2000).

\bibitem{sachdev} S. Sachdev, {\it Quantum Phase Transitions}, Cambridge University Press, Cambridge (1999).

\bibitem{affleck}
I. Affleck, Phys. Rev. B {\bf 43}, 3215 (1991).
% Bose condensation in quasi-one-dimensional antiferromagnets in strong fields

\bibitem{nikuni} T. Nikuni, M. Oshikawa, A. Oosawa, and H. Tanaka, Phys. Rev. Lett. {\bf 84}, 5868 (2000).

\bibitem{giamarchi}  T. Giamarchi, A. M. Tsvelik,
Phys. Rev. B {\bf 59}, 11398 (1999).
%Coupled ladders in a magnetic field

\bibitem{bruce}
M. Matsumoto, B. Normand, T. M. Rice und M. Sigrist,
Phys. Rev. Lett. {\bf 89}, 077203 (2002)

\bibitem{haas}
%    Title: Quantum Phase Transitions in Coupled Dimer Compounds
O. Nohadani, S. Wessel, S. Haas, preprint cond-mat/0411599.

\bibitem{rosch:2000} A. Rosch, Phys. Rev. B {\bf 62}, 4945 (2000).

\bibitem{gruen} L. Zhu, M. Garst, A. Rosch, and Q. Si, Phys. Rev. Lett. {\bf 91}, 066404 (2003).

\bibitem{hlubina} R. Hlubina and T. M. Rice, Phys. Rev. B {\bf 51}, 9253 (1995).

\bibitem{2kf}    B. L. Altshuler,  L. B. Ioffe, A. J. Millis, Phys. Rev. B 
{\bf 52}, 5563 (1995).
%CriticalbehavioroftheT=0 2kF density-wave phase transition in a two-dimensional Fermi liquid

\bibitem{luethi} We thank B. Luethi for pointing out that the
  sound velocity is an interesting quantity to study at a quantum
  critical point.


\bibitem{markusDA} M. Garst, {\it Aspects of Quantum Phase Transitions: Gr\"uneisen Parameter, Dimensional Crossover and Coupled Impurities}, Ph.D. thesis, Universit\"at Karlsruhe, 2003.



\end{thebibliography}
\end{document}